# Calculating the Exact Pooled Variance

J. W. Rudmin, Physics Dept., James Madison University
Harrisonburg, VA 22807

**ABSTRACT:** An exact method of calculating the variance of a pooled data set is presented. Its major advantages over the many other methods are that it is simple, is easily derived and remembered, and requires no assumptions. The result can be concisely summarized as follows: "The exact pooled variance is the mean of the variances plus the variance of the means of the component data sets." The proof is so simple that it has certainly been done many times before, but it is absent in the textbooks. Its practical significance is discussed.
-------------------------

## I. Introduction

Many introductory statistics textbooks deal with the fundamental principles of measurement, and this paper assumes that the reader is familiar with the different types of experimental errors and how they are handled, and with widely used notations and terminology. This paper will deal with the statistical errors, and will use the dictionary definition of variance and standard deviation: *The standard deviation* is the square root of the arithmetic average of the squares of the deviations from the mean, and the *variance* is the square of the standard deviation. The *standard error* is the estimate of how far the mean value is likely to be in error due to random errors and a finite number of measurements. This paper will use the letters i, j, k, m, & n for integers, x for data values, a for averages or means, v for variances, and S for sums of squares of the data values. Thus for a data set
$D = \{x_1, x_2 \ldots x_N\}$, N is the number of data points, the mean is $a = (x_1 + x_2 + \ldots + x_N)/N$, the variance is
$v = ((x_1-a)^2 + (x_2-a)^2 + \ldots + (x_N-a)^2)/N$, the standard deviation is $\sigma_N = v^{1/2}$, and the standard error is $SE = \sigma_N /(N-1)^{1/2}$.

Many texts use N-1 for the denominator in the variance. In this case, the standard deviation is denoted by $\sigma_{N-1}$ and the standard error is $SE = \sigma_{N-1}/N^{1/2}$. Note that SE is the same either way. SE is sometimes called the "standard deviation of the means". (See the References below for a sampling of choice of $\sigma_N$ or $\sigma_{N-1}$ )

The standard deviation is a measure of the spread of the data about the mean. It is roughly the typical distance of a data point from the mean value of the data points, but varies with the distribution of the data points around the mean. For a large number N of data points, it is nearly independent of N. The standard error is an estimate of the statistical uncertainty in the mean value of the data, and for large N works quite well for symmetric distributions of the data points around the mean. It decreases with increasing N like $N^{-1/2}$.

This reduction in SE provides a motive for pooling, or merging, data sets. More points yield a smaller statistical uncertainty in the mean. However, different experiments may actually be measuring different things. For example, in measuring the acceleration of gravity in Paris using a pendulum, one experiment may eliminate the buoyancy of the air on the bob, and another may eliminate the thermal expansion of the length of the supporting wire, and a third may hold the amplitude of the swing to be a known constant. In this case the measurements differ, and it may be invalid to pool the data from the different experiments. These are examples of consistent, rather than statistical, errors in an experiment. Sometimes these exist, but are unknown. How variance changes as more data sets are included can yield a criterion for not pooling the data sets. This paper will present a simple, exact, easily-understood formula for the variance of a pooled data set when for each subset the number of points, the mean, and the variance are known. This calculation is extremely simple, yet I have not found it in any statistics books. The author requests that this method be referred to as "the exact pooled variance".

**II. Theorem:** Let several subsets of measurements of variable be merged to form a pooled data set. For each set only the number of data points, the mean value, and the variance of each subset are given. The variance of the pooled set is the variance of the means plus the mean of the variances of the subsets. In this discussion, a "mean" is a weighted average over the subsets, and the weights are the number of data points in each subset.

In rigorous form, this can be proven for two data sets, and then extended to any number of sets by mathematical induction. In this case I will simply show it for three data sets, since I think it is clearer to do so.

Let $x_i$ represent measurements of some variable. Let $D_P = \{x_1, x_2, ... x_n\}$, be a pooled data set assembled from data sets $D_1 = \{x_1, x_2, ... x_j\}$, $D_2 = \{x_{j+1}, x_{j+2}, ... x_{j+k}\}$, and $D_3 = \{x_{j+k+1}, x_{i+j+2}, ... x_{j+k+m}\}$. Note that the number of points in the pooled set is $n = j+k+m$.

The <u>mean</u> or <u>average</u> of $D_P$ is $a_P = \dfrac{1}{n}\sum_{i=1}^{n} x_i$. Means $a_1$, $a_2$, and $a_3$ for $D_1$, $D_2$, and $D_3$, are found similarly. Then

$$a_P = (ja_1 + ka_2 + ma_3)/n = \text{the \underline{mean of the means}}. \tag{1}$$

The <u>sum of squares</u> of $D_P$ is $S_P = \sum_{i=1}^{n} x_i^2$. Sums $S_1$, $S_2$, and $S_3$ are similarly defined for sets $D_1, D_2$, and $D_3$.

Then
$$S_P = S_1 + S_2 + S_3 \tag{2}$$

Define the <u>variance</u> of data set $D_P$ to be $v_P = \dfrac{1}{n}\sum_{i=1}^{n} (x_i - a_P)^2$. Expanding the argument easily shows that
$$v_P = S_P/n - a_P^2. \tag{3}$$

Similarly, $\quad v_1 = S_1/j - a_1^2, \qquad v_2 = S_2/k - a_2^2, \text{ and } \qquad v_3 = S_3/m - a_3^2. \tag{4}$

Our goal is to calculate $v_P$ from the means and variances of the constituent sets.

From (2), (3), and (4), $\quad nv_P = S_1 + S_2 + S_3 - na_P^2 = j(v_1+a_1^2) + k(v_2+a_2^2) + m(v_3+a_3^2) - na_P^2$.

Or, $\quad nv_P = jv_1 + kv_2 + mv_3 + ja_1^2 + ka_2^2 + ma_3^2 - na_P^2. \tag{5}$

The <u>mean of the variances</u> is $\qquad a(v) = (jv_1+kv_2+mv_3)/n. \tag{6}$

From (1), the <u>variance of the means</u> is $\quad v(a) = [j(a_1-a_P)^2 + k(a_2-a_P)^2 + m(a_3-a_P)^2]/n$

or $\quad nv(a) = ja_1^2 + ka_2^2 + ma_3^2 - 2a_P(ja_1 + ka_2 + ma_3) + na_P^2 = ja_1^2 + ka_2^2 + ma_3^2 - na_P^2 \tag{7}$

Combining (1), (5), (6) and (7) yields $\qquad v_P = a(v) + v(a) \quad$ QED $\tag{8}$

**Thus the variance of the pooled set is the mean of the variances plus the variance of the means**.

---------------------------------------------------------------------------

### III. A Numerical Example

Consider three data sets D1, D2, and D3, which are pooled to create data set Dp.

Data set D1 contains 6 points:  32  36  27  28  30  31
Data set D2 contains 7 points:  32  34  30  33  29  36  24
Data set D3 contains 3 points:  39  40  42

The pooled set Dp contains 16 points:  32  36  27  28  30  31  32  34  30  33  29  36  24  39  40  42

Now compute the mean and variance for each data set using the data points.
D1:  Mean = 30.66667          Variance = 8.555555
D2:  Mean = 31.14286          Variance = 13.26531
D3:  Mean = 40.33333          Variance = 1.555555
Dp:  Mean = 32.6875           Variance = 22.83984

The mean of the variances  = 9.303571
The variance of the means  = 13.53627
Then the exact pooled variance = 22.83984     Compare this with the variance of Dp above.

Also note how adding D3 to D1 and D2 hardly changes the mean of the data, but doubles the variance. D3 has a much smaller variance than D1 or D2, so including it using precision-weighted adjustment procedures could drastically change the mean and variance of the combined adjusted data. These will be discussed later.

### IV. A Simple Physical Interpretation

A colleague noted that this is the parallel-axis theorem of classical mechanics. Consider a structure such as a tree-branch with apples on it. The mean value of the positions of the masses is the center of mass, and the moment of inertia about the center of mass is just the variance of the distance of the masses from the center of mass. The parallel-axis theorem was proven by Jakob Steiner (1796-1863) in two dimensions. This theorem uses only one dimension. The book Statistics, Probability, and Decisions by Winkler also mentions the connection between variance and moment of inertia, so in a loose sense, this may be the earliest "proof" of the above theorem.

### V. Advantages of Using the Exact Pooled Variance instead of Using the Mean Variance as an Estimate

Many introductory statistics texts recommend using the mean of the variances as an estimate of the variances of the pooled set. This practice should be abandoned for the following reasons.
1. This practice will always yield a pooled variance which is smaller than the exact pooled variance.
2. The exact variance is easily calculated, and the data to do so is always available in a pooling process.
3. It is unreasonable to include the caution "if the means of the pooled sets are approximately equal", when no objective criterion is given for "approximately equal" and the easiest objective one is simply calculating the variance of the means--the second term in the exact pooled variance, which removes the need for approximating.
4. In pooling data sets, if the mean, standard deviation, and standard error are to have any reasonably valid meaning, the pooled variance must not ignore the difference of the means of the constituent sets. If two data sets of a time measurement are pooled, and one set has mean $a_1 = 10s$, and standard deviation $\sigma_1=3s$, and the other has mean $a_2=20s$, and $\sigma_2=3s$, then to say that the pooled set has mean $a_P=15s$ and $\sigma_P=2s$, is grossly misleading.
5. If adding more data points to the pool increases the standard error, then this can be a signal that the measurement methods need examination, and that perhaps the new data should not be included.
6. If the input data sets do not use adjusted data, then the Exact pooled variance is well-defined, associative, and commutative. More data sets can be added into a pooled variance, and the result is the same, independent of the order in which the data sets are accumulated.
7. The exact pooled variance is independent of the distribution of the data points and requires no assumptions that any properties of the data distribution be negligible.
8. In contrast to most other methods, the exact pooled variance clearly teaches the principles of statistical analysis of data, and can validly be used to evaluate results.
9. The procedure for calculating the exact pooled variance can be concisely stated, and is easily memorized.

### VI. Can a Decrease in Standard Error be a Criterion for the Validity of Pooling?

This brings us to an obvious question: If the goal of pooling is to decrease the standard error, then why pool if pooling increases the standard error? Or suppose you already have a pooled data set, and are considering adding a new data set to the pool (which is easily done for the exact pooled variance). Then suppose that including this new data set increases the variance. Would this be a valid justification for excluding the new set?

It would clearly valid *if* there are other reasons to suspect a flawed measurement. For example, suppose thirty students in a physics lab each measure e/m, the charge-to-mass ratio of the electron. The lab has, say ten commercially-manufactured instruments, all of the same make and model. The instructor leads the class in pooling the data to show how increasing the number of measurements decreases the standard error. He then discovers that three of the measurements are a factor of two lower than the rest of the measurements, and all three are from instrument #1. The other nine give e/m = 0.36 ± 0.03 TC/kg, and #1 gives e/m = 0.18 ± 0.05 TC/kg. He quickly checks instrument #1 himself and measures 0.179 TC/kg, so he believes that the students using #1 were following the correct procedure. In this case it would be reasonable to assume that #1 is doing something different which halves its results compared to the other instruments and he excludes its results from the pool. The instructor also notes that instrument #1 is consistent with the internationally accepted best value for e/m. He then checks the other instruments and then notifies the manufacturer that nine of the ten instruments are consistently giving a result which is twice the CODATA[14] accepted value. Notice that in this example, the increase in the variance was grounds for rejecting instrument #1 from the pool, allowing a consistency for the other nine instruments which suggested a possible flaw in their design or construction.

A disagreement between two measurements does not indicate which of the measurements is more accurate. However it may indicate that further examination of the instruments is warranted, and in some cases may lead to new, unexpected information. In some cases, both measurements might be published, but not combined into a common pool.

**VII. Adjustment of Data and Analysis of Variance.**

A method of combining data sets which differs from simple pooling is widely-used in science, and is called "adjustment of data". This procedure is used by the CODATA committee to determine a best value for fundamental constants.[10,14,] This method seeks to generate a weighted average of mean values from different sources. Usually the weights are chosen using the variance, standard deviation, or chi-squared functions of the composite data sets, and are chosen to minimize at least one of these functions. Suffice it to say, this procedure is not a pooling of data sets, but rather a combining of data sets which does not incorporate the differences of the mean values of the composite sets. This subject is not further discussed here, so as not to distract from the main point of the paper.

"Analysis of Variance" (ANOVA) comprises a vast literature dating back to at least 1932[12], and summaries of this method are available in statistics textbooks[13]. In ANOVA, both of the terms a(v) and v(a) in equation (8) above are calculated, but instead of adding them to give the exact pooled variance, the ratio v(a)/a(v) is considered as a correction to the mean of the variances. This ratio is then used with various criteria to try to achieve various goals, such as comparing effects of different medical treatments, maximizing profits, improving quality control in manufacturing, or minimizing catastrophic failures. In any of the numerous discussions of ANOVA it would have been trivial to include the exact pooled variance, but I have not yet found a case of this being done.

**VIII. Conclusions**

1. The exact variance of a data set formed by pooling subsets is the mean of the variances plus the variance of the means of the subsets. The means are weighted by the number of points in each subset.

2. This method of calculating the variance of pooled data sets is probably not original, even though it is not widely known. In the sense of being a special case of the parallel axes theorem, it has been done before, but is not applied in statistics books and journals. The author would appreciate hearing of other earlier publications of this method.

3. Advantages of using the exact pooled variance include that it is simple, requires no assumptions, is exact, is easily understood and remembered, is pedagogically helpful, and can yield evidence for possible consistent errors suggesting rejection of the pooling and/or investigation of the measurements. Because of these advantages, the calculation of the exact pooled variance should be included in every introductory statistics textbook.